\documentclass[final]{aa}

\usepackage[dvips]{graphicx}
\usepackage{epsfig}
\usepackage{txfonts}
\usepackage{longtable}
\usepackage{color}
\usepackage{soul}
\usepackage{natbib}
\usepackage{hyperref}
\hypersetup{colorlinks=true, urlcolor=blue}
\usepackage{booktabs}
\usepackage{multirow}
\usepackage{mdwtab}
\usepackage{array}
\usepackage{makecell}

\begin{document}

   \title{Detection of planet candidates around K giants \thanks{Based on observations made with the BOES at BOAO in Korea and HIDES at OAO in Japan.} \\HD 40956, HD 111591, and HD 113996}
   \author{G. Jeong\inst{1,2},
                B.$-$C. Lee\inst{1,2} \thanks{Corresponding author : B. -C. Lee},
                I. Han\inst{1,2},
                M. Omiya\inst{3},
                H. Izumiura\inst{3,4}
                B. Sato\inst{5},
                H. Harakawa\inst{3},
                E. Kambe\inst{4}
                \and D. Mkrtichian \inst{6}
          }

   \institute{Korea Astronomy and Space Science Institute, 776,
                Daedeokdae-Ro, Youseong-Gu, Daejeon 34055, Korea\\
              \email{[tlotv;bclee;iwhan]@kasi.re.kr}
            \and
                Korea University of Science and Technology,
        217, Gajeong-ro, Yuseong-gu, Daejeon 34113, Korea
           \and
                 National Astronomical Observatory of Japan, 2-21-1 Osawa, Mitaka, Tokyo 181-8588, Japan\\
                \email{[omiya.masashi;h.harakawa]@nao.ac.jp}
           \and
                Okayama Astrophysical Observatory, National Astronomical Observatory of Japan, 3037-5 Honjo, Kamogata, Asakuchi, Okayama 719-0232, Japan\\
                \email{[izumiura;kambe]@oao.nao.ac.jp}
           \and
                Department of Earth and Planetary Sciences, Tokyo Institute of Technology, 2-12-1 Okayama, Meguro-ku, Tokyo 152-8551, Japan\\
                \email{satobn@geo.titech.ac.jp}
           \and
                National Astronomical Research Institute of Thailand, Siriphanich Bldg., Chiang Mai, Thailand\\
                \email{davidmkrt@gmail.com}
             }

   \date{Received 28 June 2016 / Accepted 18 October 2017}

  \abstract
   {}
   {
   The purpose of this paper is to detect and investigate the nature of long-term radial velocity (RV) variations of K-type giants and to confirm planetary companions around the stars. 
   }
   {
   We have conducted two planet search programs by precise RV measurement using the 1.8 m telescope at Bohyunsan Optical Astronomy Observatory (BOAO) and the 1.88 m telescope at Okayama Astrophysical Observatory (OAO). The BOAO program searches for planets around 55 early K giants. The OAO program is looking for 190 G--K type giants.
   }
   {
   In this paper, we report the detection of long-period RV variations of three K giant stars, \mbox{HD 40956}, \mbox{HD111591}, and \mbox{HD113996}.
   We investigated the cause of the observed RV variations and conclude the substellar companions are most likely the cause of the RV variations.
   The orbital analyses yield P = 578.6 $\pm$ 3.3 d, $m$ sin $i$ = 2.7 $\pm$ 0.6 $M_{\rm{J}}$, $a$ = 1.4 $\pm$ 0.1 AU for \mbox{HD 40956}; P = 1056.4 $\pm$ 14.3 d, $m$ sin $i$ = 4.4 $\pm$ 0.4 $M_{\rm{J}}$, $a$ = 2.5 $\pm$ 0.1 AU for \mbox{HD 111591}; P = 610.2 $\pm$ 3.8 d, $m$ sin $i$ = 6.3 $\pm$ 1.0 $M_{\rm{J}}$, $a$ = 1.6 $\pm$ 0.1 AU for \mbox{HD113996}.
   }
   {}


   \authorrunning{G. Jeong et al.}
   \titlerunning{Detection of planetary companions around K giants}
   \maketitle
%
\section{Introduction}
Planetary formation around normal stars involves complex processes that depend on various stellar properties: stellar mass, metallicity, radiation flow of central stars, the host star's environment, binarity, and so on. 
An understanding of the planetary formation can be done only if the effects of these stellar parameters are disentangled.
To do this it is necessary to search for planets around stars with a wide range of stellar properties (especially stellar mass).
However, the majority of exoplanets have been detected around solar-type main sequence stars by radial velocity (RV) and transit methods.

Fifteen years ago, we had only limited knowledge about exoplanets around intermediate-mass stars with stellar masses of more than 1.5 $M_{\odot}$. The intermediate-mass main sequence stars are not suitable targets for precise RV measurements because they have few absorption lines that are often broadened due to high stellar-rotation rates.
Although some precise Doppler surveys searching for planets around intermediate-mass stars had been conducted, we didn't know how each process in the planetary formation depends on the stellar mass larger than 1.5 $M_{\odot}$.
However, recently, the Doppler surveys of G- and K-type giant stars located in the red clump on the H-R diagram have uncovered the properties of planetary systems around intermediate-mass stars because the giants are evolved intermediate-mass stars with slower rotations and lower temperatures suitable for precise RV measurements (e.g., \citealt{2012PASJ...64..135S}).

Beginning with the first hints of possible planets around K giant stars ($\beta$ Gem) found by \citet{1993ApJ...413..339H}, the discovery of the giant planet around $\iota$ Dra by \citet{2002ApJ...576..478F} is largely considered as the first confirmed planet around a giant star. The number of planets orbiting giant stars is increasing and more than 95 planets have been detected so far. Stellar properties of the planet-hosting giants have been understood with a significantly large sample. 
Based on a Doppler survey of 373 G- and K-type giants, \citet{2015A&A...574A.116R} found that the planet occurrence rate of intermediate-mass stars depends on stellar masses, and the rate reaches its maximum at a stellar mass of 1.9 $M_{\odot}$ and drops off at stellar masses of more than 2.7 $M_{\odot}$. 
They concluded that no giant planet has been found around stars more massive than 2.7 $M_{\odot}$ and suggested that the giant planet formation and/or inward migration in protoplanetary disks are suppressed for more massive stars because of fast disk depletion and/or the long migration time scale of giant planets.

The photometric transit method may be more favorable for Jupiter-type exoplanet detection around rapidly rotating early-type stars. However, only five planets have been found around F- and A-type stars; WASP-33 (\citealt{2010MNRAS.407..507C}),  KOI\,13 (\citealt{2011ApJS..197....2F}), HAT-P-57 (\citealt{2015AJ....150..197H}), HAT-P-67 (\citealt{2017AJ....153..211Z}), KELT-17 (\citealt{2016AJ....152..136Z}) and KELT-9b (\citealt{2017Natur.546..514G}), and some of them have been confirmed by the spectroscopic transiting method using Doppler Tomography. Thus, the most significant planet searches for giant planets around intermediate-mass stars are precise Doppler surveys of evolved giant stars.

Since 2003, we have been carrying out a precise RV survey of 55 K0--K4-type giants to investigate RV variations of K giant stars and to search for planets around the stars (\citealt{2008JKAS...41...59H,2010A&A...509A..24H}) using the 1.8 m telescope at Bohyunsan Optical Astronomy Observatory (BOAO) and the Bohyunsan Observatory Echelle Spectrograph (BOES). To date we have discovered nine planet candidates around K giant stars (e.g., \citealt{2014A&A...566A..67L}).

The East-Asian Planet Search Network, EAPS-Net(\citealt{2005JKAS...38...81I}) is an international collaboration to search for planets around intermediate-mass G- and K-type (sub)giants using the 2 m-class telescopes in China, Japan, and Korea. In the framework of the EAPS-Net, the Korean-Japanese planet search program has been performed using the 1.8 m telescope at BOAO and BOES in Korea and the 1.88 m telescope at Okayama Astrophysical Observatory (OAO) and the HIgh Dispersion Echelle Spectrograph (HIDES) in Japan (\citealt{2009PASJ...61..825O, 2012PASJ...64...34O}). In the program, 80 and 110 giants have been monitored at BOAO and OAO, respectively. We have also made intensive follow-up observations for candidate stars at both observatories to determine orbital parameters. 

This paper is organized as follows. In Section 2, we describe the RV observations using the BOAO and OAO telescopes. Stellar properties and orbital solution of the planets are presented in Sections 3 and 4. In Section 5, we investigate the cause of radial velocity variations by \emph{HIPPARCOS} photometry, chromospheric activity, and line shape analysis. Section 6 contains the summary and conclusion of this paper.

\section{RV Observations}
\subsection{BOES observations}
At BOAO, we have performed spectroscopic observations using the 1.8 m telescope and the Bohyunsan Observatory Echelle Spectrograph (BOES; \citealt{2007PASP..119.1052K}), which is a fiber-fed high-resolution echelle spectrograph. Using the BOES, we collected 22, 24, and 64 data points for \mbox{HD 40956}, \mbox{HD 111591}, and \mbox{HD 113996}, respectively. The data of HD 40956 and HD 111591 were obtained with the 200 $\mu$m fiber, giving a spectral resolution of 45,000. The typical exposure time of the observations is around 20 minutes, which can achieve a signal to noise ratio (S/N) of more than 150. HD 113996 was observed by the 80 $\mu$m fiber, giving a spectral resolution of 90,000. The average exposure time was 8 minutes to get a S/N of about 150.

One-dimensional (1D) spectra were extracted from the two-dimensional (2D) raw frames using standard Image Reduction and Analysis Facility (IRAF) procedures for bias subtraction, flat fielding, scattered light correction, spectrum extraction, and ThAr wavelength calibration. Precise RV measurements were performed by the program RVI2CELL (\citealt{2007PKAS...22...75H}). The long-term RV accuracy of BOES is about 7 m s$^{-1}$ which is estimated by long-term monitoring of a RV standard star $\tau$ Ceti (\citealt{2013A&A...549A...2L}). We also use a pure high-resolution spectrum taken without the I$_{2}$ cell by BOES to determine stellar parameters.

\subsection{HIDES observations}
We also made RV observations of HD 40956 and HD 111591 in the framework of the Korean-Japanese planet search program using the 1.88 m telescope and the HIgh Dispersion Echelle Spectrograph (HIDES, \citealt{1999PYunO....S..77I}, \citealt{2013PASJ...65...15K}) at OAO. In December 2007, HIDES CCD system was upgraded from a single 2k x 4k CCD to a mosaic CCD system (three 2k x 4k CCDs) covering a wide wavelength range of 3800--7500 \AA, which is three times larger than before. Although we made the observations using a slit mode of the HIDES until 2011 (HIDES-slit), HIDES was also upgraded to add a fiber-fed mode using a fiber-fed system attached to the Cassegrein focus of the telescope (HIDES-fiber). For precise RV measurements of our targets, we used an I$_{2}$ cell put in front of the slit and the fiber entrance for HIDES-slit and HIDES-fiber modes, respectively (\citealt{2002PASJ...54..865K}). We set a slit width at 0.76 arcsec, giving a spectral resolution of 63,000 when we used the HIDES-slit mode. Using the 100 $\mu$m fiber of the fiber-fed mode, we get a spectral resolution of 50,000. Typical exposure times of the observations of HD 40956 and HD 111591 at OAO are 25 and 15 minutes with HIDES-slit and HIDES-fiber modes, respectively, achieving S/Ns of about 170.

Reduction of the HIDES data was done using software packages of the IRAF.
The standard reductions of bias subtraction, flat-fielding, order extraction, wavelength calibration,  and so on, were performed on each frame.
Stellar RV measurements were performed with a modeling technique using an I$_{2}$ superposed stellar spectrum detailed in \cite{2002PASJ...54..873S} based on a method of \cite{1996PASP..108..500B}. Stellar templates were made by deconvolutions from pure stellar spectra taken without the I$_{2}$ cell with instrumental profiles estimated from I$_{2}$ superposed flat spectra (\citealt{2007ApJ...661..527S}). Typical errors of the RV measurements are about 3.5--6 m s$^{-1}$. We obtained 36 and 20 spectra for HD 40956 and HD 111591, respectively.

The RVs of each star derived from BOES and HIDES data are listed in Tables ~\ref{tab3},~\ref{tab4}, and ~\ref{tab5}.

\begin{table*} [h]
\begin{center}
\caption[]{Stellar parameters for the stars in this study.}
\label{tab1}                             
\begin{tabular}{ccccc}
\toprule[1.5pt]

    Parameter        & HD 40956   & HD 111591   & HD 113996   & Ref.     \\

\midrule
    Spectral type           & K0   & K0 III   & K5 III   & 1     \\
    $\textit{$m_{v}$}$ (mag) & 6.584 $\pm$ 0.001   & 6.594 $\pm$ 0.001   & 4.918 $\pm$ 0.001   & 1  \\
    $\textit{B-V}$     (mag) & 1.011 $\pm$ 0.006  & 1.002 $\pm$ 0.002  & 1.482 $\pm$ 0.003  & 1 \\
    RV                 (km s$^{-1}$) & $-$ 15.80 $\pm$ 0.25 & 5.67 $\pm$ 0.17 & $-$ 15.72 $\pm$ 0.01 &  2 \\
    $\pi$          (mas) & 8.44 $\pm$ 0.51  & 9.22 $\pm$ 0.47  & 9.84 $\pm$ 0.22 & 1 \\
    $T_{\rm{eff}}$     (K)& 4869 $\pm$ 28 & 4884 $\pm$ 30 & 4181 $\pm$ 40 &  3 \\
    $\rm{[Fe/H]}$      & 0.14 $\pm$ 0.05   & 0.07 $\pm$ 0.04   & 0.13 $\pm$ 0.08  & 3  \\
    log $\it g$        (cgs) & 3.02 $\pm$ 0.09  & 3.10 $\pm$ 0.10  & 1.86 $\pm$ 0.16  &  3  \\
    $\varv_{\rm{micro}}$ (km s$^{-1}$)       & 1.20 $\pm$ 0.10  & 1.23 $\pm$ 0.11  & 1.53 $\pm$ 0.14  & 3  \\
   Age (Gyr)        & 1.35 $\pm$ 0.18      & 1.41 $\pm$ 0.14    & 3.24 $\pm$ 1.20 & 3 \\
    $\textit{$R_{\odot}$}$ ($R_{\odot}$) & 8.56 $\pm$ 0.33  & 8.03 $\pm$ 0.49  & 25.11 $\pm$ 1.20  & 3 \\
    $\textit{$M_{\odot}$}$ ($M_{\odot}$) & 2.00 $\pm$ 0.08  & 1.94 $\pm$ 0.07  & 1.49 $\pm$ 0.18  & 3 \\
    $\textit{$L_{\odot}$}$ ($L_{\odot}$) & 46.17  & 38.07   & 291.00   &  2   \\
    $\varv_{\rm{rot}}$ sin $i$ (km s$^{-1}$) & 2.7 $\pm$ 0.5 & 3.1 $\pm$ 0.5    & 3.3 $\pm$ 0.5   & 3 \\
    $P_{\rm{rot}}$ / sin $i$ (days)  & 158.9  & 130.5  & 389.9  & 3 \\

\bottomrule[1.5pt]

\end{tabular}
\end{center}
\textbf{References.}--- (1) van Leeuwen (2007); (2) Anderson \& Francis (2012); (3) This work. \\
\end{table*}

\begin{table*}
\begin{center}
\caption[]{Orbital parameters for the companions.}
\label{tab2}
\begin{tabular}{lccc}
\toprule[1.5pt]

    Parameter        & HD 40956 b   & HD 111591 b   & HD 113996 b  \\
\midrule
    P (days)           & 578.6 $\pm$ 3.3   & 1056.4 $\pm$ 14.3   & 610.2 $\pm$ 3.8   \\
    K (m s$^{-1}$)   & 68 $\pm$ 2   & 59 $\pm$ 3  & 120 $\pm$ 9   \\
    $T_{periastron}$ (JD)  & 2455341.88 $\pm$ 11.07 & 2455602.40 $\pm$ 30.85 & 2453309.94 $\pm$ 23.67  \\
    $e$                    & 0.24 $\pm$ 0.05  & 0.26 $\pm$ 0.10  & 0.28 $\pm$ 0.12  \\
    $\omega$ (deg)  & 338.62 $\pm$ 7.74 & 78.72 $\pm$ 11.94 & 92.06 $\pm$ 17.25  \\
    $m$ sin $i$ ($M_{\rm{J}}$) & 2.7 $\pm$ 0.6 & 4.4 $\pm$ 0.4 & 6.3 $\pm$ 1.0 \\
    $a$ (AU)           & 1.4 $\pm$ 0.1  & 2.5 $\pm$ 0.1  & 1.6 $\pm$ 0.1  \\
    Slope (m s$^{-1} yr^{-1}$) & -13 $\times$ 10$^{-2}$  & --  & -- \\
    $N_{obs}$        & 54 & 44 & 62  \\
    rms (m s$^{-1}$) (BOAO) & 23.6  & 19.5  & 39.3\\
    rms (m s$^{-1}$) (OAO) & 16.1  & 9.9  & \\
\bottomrule[1.5pt]

\end{tabular}
\end{center}
\end{table*}
\section{Stellar properties}
The stellar properties of HD 40956, HD 111591, and HD 113996 are summarized in Table~\ref{tab1}. Visual magnitude ($m_{v}$) and parallax ($\pi$) were adopted from \citet{2007A&A...474..653V} revised from the \emph{HIPPARCOS} catalog (\citealt{1997yCat.1239....0E}). The atmospheric parameters (effective temperature $T_{\rm{eff}}$, Fe abundance [Fe/H], surface gravity log $\it g$, and microturbulent velocity $\varv_{\rm{micro}}$) were derived by the TGVIT (\citealt{2005PASJ...57...27T}) program, which uses equivalent width (EW) of Fe I and Fe II lines.
The projected rotational velocity $\varv_{\rm{rot}}$ sin $i$ was estimated by \citet{2008PASJ...60..781T}.
The spectral type, RV, and luminosity of these stars were obtained from \citet{2012AstL...38..331A}. The stellar radius, mass, and age of each star were calculated by the method of \citet{2006A&A...458..609D}.

HD 40956 is a K0-type star with an effective temperature of 4869 $\pm$ 28 K, a stellar luminosity of 46.17 $L_{\odot}$, and a metallicity [Fe/H] of 0.14 $\pm$ 0.05. It seems to be located on the red clump on the H-R diagram.

HD 111591 is a K0 III-type giant star with an effective temperature of 4884 $\pm$ 30 K, a stellar luminosity of 38.07 $L_{\odot}$, and a metallicity [Fe/H] of 0.07 $\pm$ 0.04.

HD 113996 is a K5 III-type giant and slightly later than the previous two stars. The star has an effective temperature of 4181 $\pm$ 40 K, a stellar luminosity of 291 $L_{\odot}$, and a metallicity [Fe/H] of 0.13 $\pm$ 0.08; it appears to be located on the RGB branch on the H-R diagram. The rotation velocity is estimated to be 3.3 $\pm$ 0.5 \mbox{km s$^{-1}$}.
%

\section{Orbital solutions}
Assuming that the RV variation is caused by Keplerian orbital motion by unseen companions, we calculated the orbital parameters using the iterated non-linear least-squares method.
During the orbital solution, we included the offset between BOES and HIDES RVs as an unknown parameter.
If the RV showed linear variation, we also included a linear slope as an unknown parameter.
The parameters of orbital solution are summarized in \mbox{Table \ref{tab2}}.

 \begin{figure} [h]
 \centering
 \includegraphics[width=8cm]{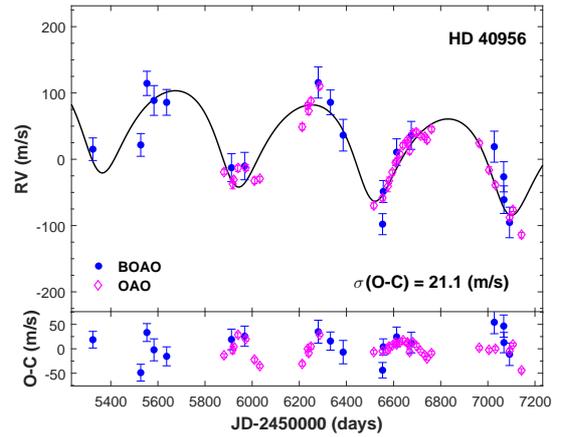}
 \caption{\emph{Top} : The RVs of HD 40956 observed at BOAO (blue dots) and OAO (magenta diamonds). A Keplerian orbital fit is shown by the solid line. \emph{Bottom} : The residuals after subtracting the orbital motion and a long-term trend from the RVs.
      }
         \label{fig1}
   \end{figure}

\begin{figure} [h]
 \centering
 \includegraphics[width=8cm]{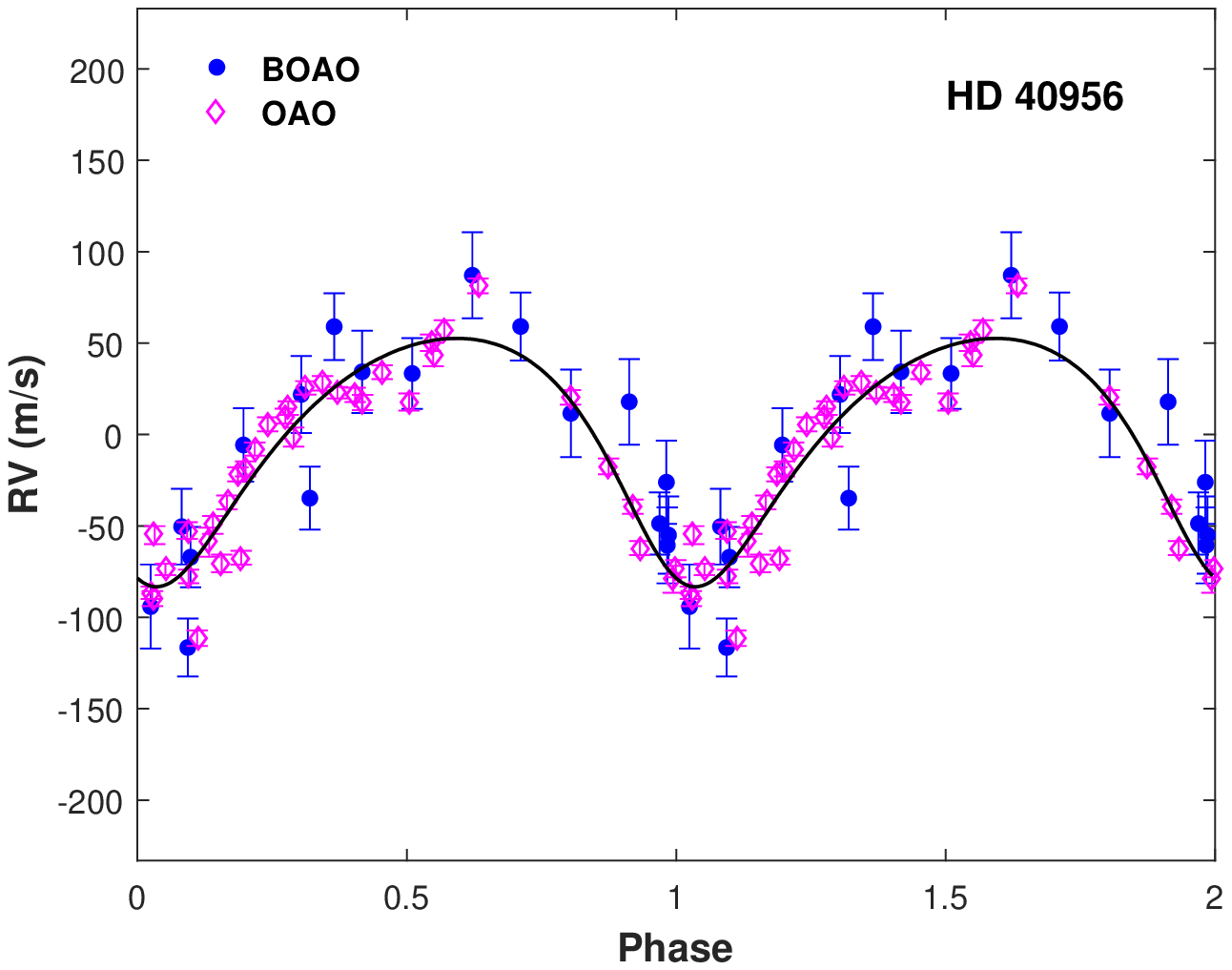}
 \caption{The phase RV curve folded by an orbital period of 578 d.
      }
         \label{fig2}
   \end{figure}
   
 \begin{figure} [h]
 \centering
 \includegraphics[width=8cm]{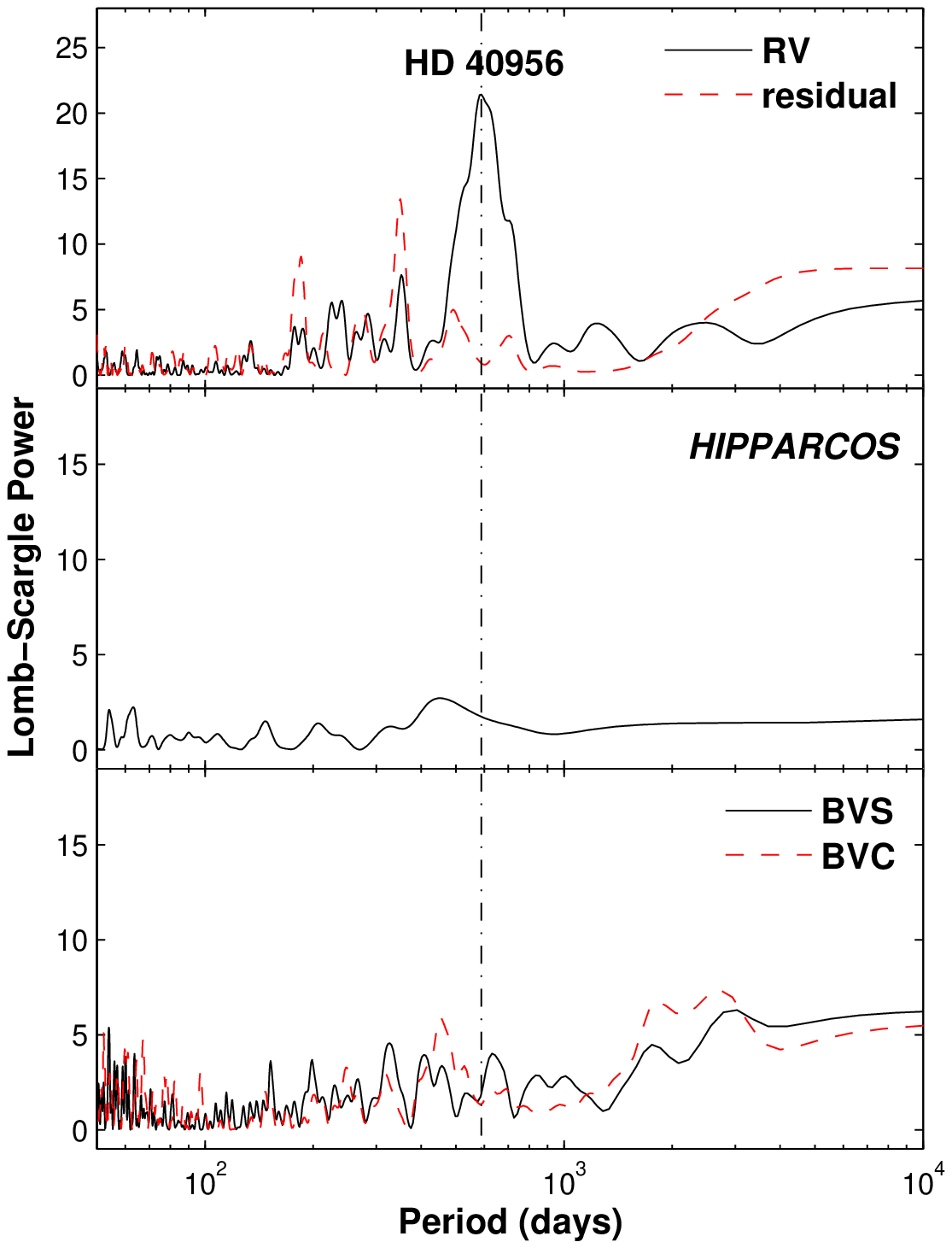}
 \caption{The L-S periodograms of the RVs and residuals (\emph{top}), \emph{HIPPARCOS} photometries (\emph{middle}), and the line bisector span (BVS, solid line) and curvature (BVC, dash line) (\emph{bottom}) for HD 40956. The vertical dash-dot line indicates an orbital period of 578 d.
    }
        \label{fig3}
 \end{figure}
 
  \begin{figure} [h]
 \centering
 \includegraphics[width=8cm]{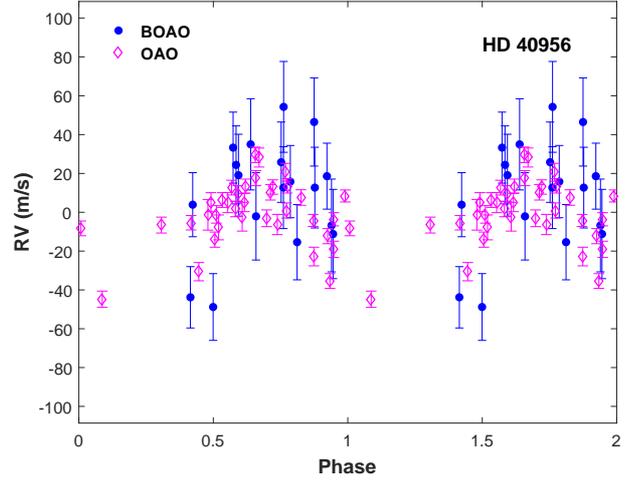}
 \caption{360-day period phase folded plot of residual RV of \mbox{HD 40956}.}
        \label{fig12}
 \end{figure}
 
\subsection{HD 40956}
The RVs of HD 40956 are listed in Table \ref{tab3} and shown in Figures \ref{fig1} and \ref{fig2}. A periodic variation is clearly seen in the RV plot. We made a Lomb-Scargle (L-S) periodogram to look for a period of the variation (Top panel of Fig. \ref{fig3}) and found a dominant peak around 578 d.
The RV also clearly shows a linear trend. Therefore, we included linear slope as an unknown parameter during the orbital solution.
The orbital fit yields an orbital period of P = \mbox{578.6 $\pm$ 3.0 d}, a semi-amplitude of $K$ = 68 $\pm$ 2 m s$^{-1}$, and an eccentricity of \mbox{$e$ = 0.24 $\pm$ 0.05}. 
Figure \ref{fig2} shows a phase folded RV of the orbital fit.
The rms of the RV residuals after the removal of the orbital motion is 21.1 m s$^{-1}$, which is somewhat larger than our RV measurement error of 12.4 m s$^{-1}$. This is not surprising, because most K giant stars show several tens m s$^{-1}$ of intrinsic RV variation (\citealt{2005PASJ...57...97S,2006A&A...454..943H}).

There seems to be small periodic variation in the residual as can be seen from the bottom panel of Figure \ref{fig1}. The result of L-S period analysis of the residual shows small peaks at 185 and 352 d (Fig. \ref{fig3}).
The 352-day period may be real or alias due to observation bias.
As can be seen from Figure \ref{fig12}, most of observations concentrate into one-half of the whole phase for 360-day periodicity. 
Therefore, we cannot determine the reality of the 352-day period at this stage. 
Adopting the stellar mass of 2.0 $M_{\odot}$, a minimum mass and a semi-major axis of the orbiting companion are estimated: $m$ sin $i$ = 2.7 $M_{\rm{J}}$, $a$ = 1.4 AU. 

\subsection{HD 111591}
\begin{figure} [h]
 \centering
 \includegraphics[width=8cm]{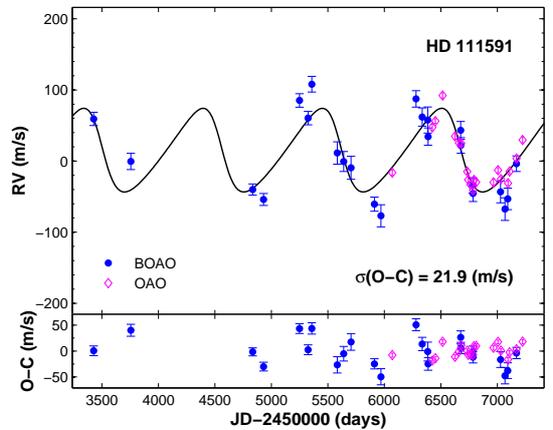}
 \caption{ \emph{Top} : The RVs of HD 111591 observed at BOAO (blue dots) and OAO (magenta diamonds). A Keplerian orbital fit is shown by the solid line. \emph{Bottom} : The RV residuals after the orbital fit.
      }
         \label{fig4}
   \end{figure}

The RVs of HD 111591 are shown in Figure \ref{fig4}. A periodic variation is clearly seen in the RV measurements. We performed a L-S period analysis and found a significant peak around 1056 d in the periodogram. The orbital fit yields a period of P = \mbox{1056.4 $\pm$ 14.3 d}, a semi-amplitude of $K$ = 59 $\pm$ 3 m s$^{-1}$, and an eccentricity of \mbox{$e$ = 0.26 $\pm$ 0.10}. The rms of the RV residuals after the removal of the 1056 d orbital motion is 21.9 m s$^{-1}$.
The residual does not show any significant periodicity (Fig. \ref{fig6}).
Adopting the stellar mass of 1.9 $M_{\odot}$, a minimum mass and a semi-major axis of the orbiting companion are estimated: $m$ sin $i$ = 4.4 $M_{\rm{J}}$, $a$ = 2.5 AU.

  \begin{figure} [h]
 \centering
 \includegraphics[width=8cm]{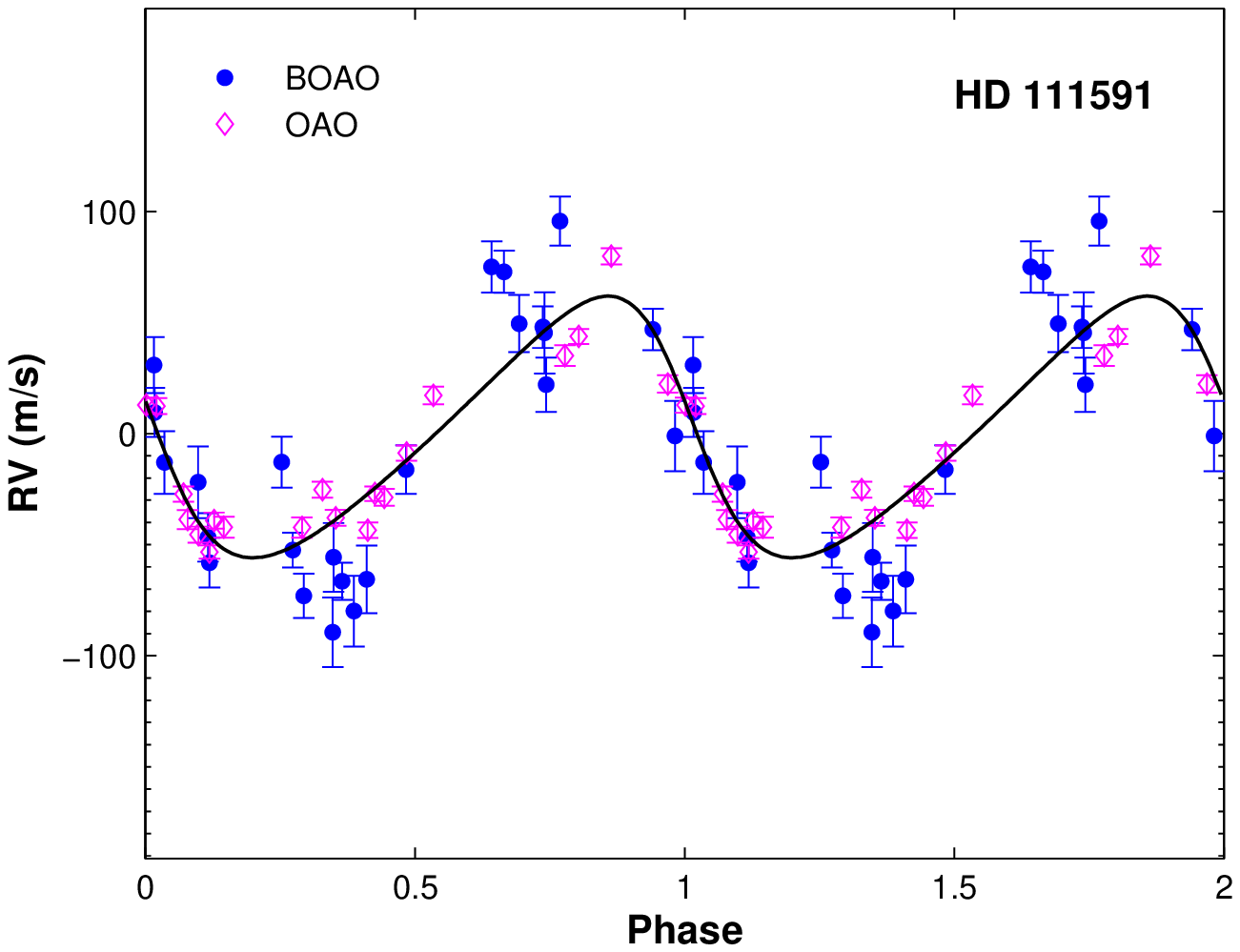}
 \caption{The phase RV curve folded by an orbital period of 1056 d.
      }
         \label{fig5}
   \end{figure}

\begin{figure} [h]
 \centering
 \includegraphics[width=8cm]{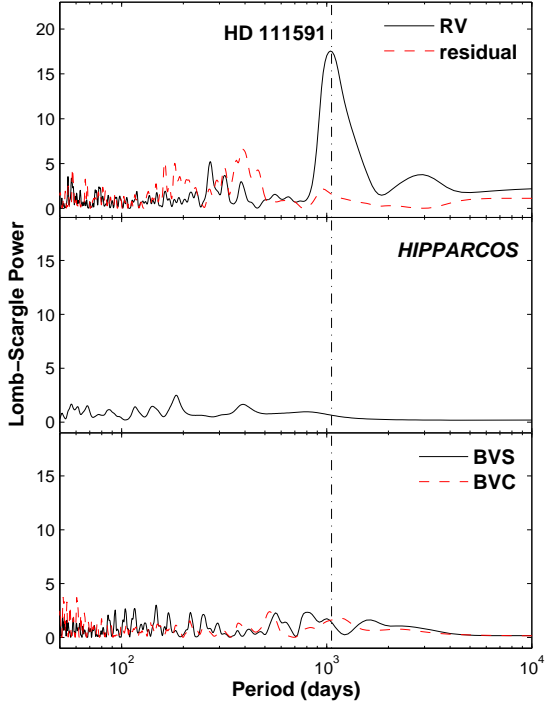}
 \caption{The L-S periodograms of the RVs and residuals (\emph{top}), \emph{HIPPARCOS} photometries (\emph{middle}), and the line bisector span (BVS, solid line) and curvature (BVC, dash line) (\emph{bottom}) for HD 111591. The vertical dash-dot line indicates an orbital period of 1056 d.
      }
         \label{fig6}
   \end{figure}

\subsection{HD 113996}
\begin{figure} [h]
 \centering
 \includegraphics[width=8cm]{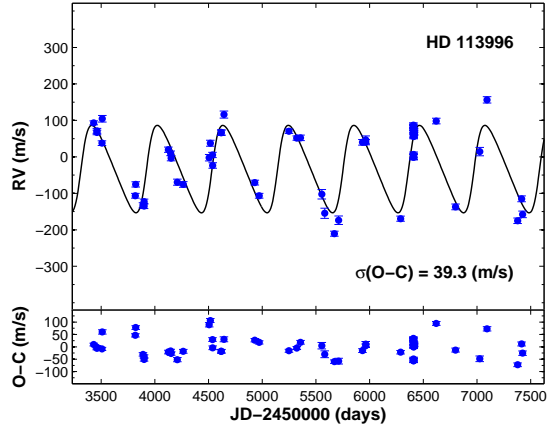}
 \caption{\emph{Top} : The RVs of HD 113996 observed at BOAO (blue dots).
 A Keplerian orbital fit is shown by the solid line. \emph{Bottom} : The RV residuals after the orbital fit.
      }
         \label{fig7}
   \end{figure}

The RVs of HD 113996 are shown in Figure \ref{fig7}. We performed a L-S period analysis and found a significant peak around 610 d. The orbital fit yields an orbital period of P = \mbox{610.2 $\pm$ 3.8 d}, a semi-amplitude of $K$ = 120 $\pm$ 9 m s$^{-1}$, and an eccentricity of \mbox{$e$ = 0.28 $\pm$ 0.12}.
The rms residual after the orbital fit is 39.3 \mbox{m s$^{-1}$}, which is larger than $~$21 m s$^{-1}$ of HD 40956 and HD 111591.
This is not surprising since HD 113996 is K5 which is later than K0 of HD 40956 and HD 111591. In general, later spectral giant stars show larger intrinsic RV variations.
The residual does not show any significant periodicity (Fig. \ref{fig9}).
Adopting a stellar mass of 1.49 $M_{\odot}$, a minimum mass and a semi-major axis of the orbiting companion are estimated: $m$ sin $i$ = 6.3 $M_{\rm{J}}$, $a$ = 1.6 AU.

 \begin{figure} [h]
 \centering
 \includegraphics[width=8cm]{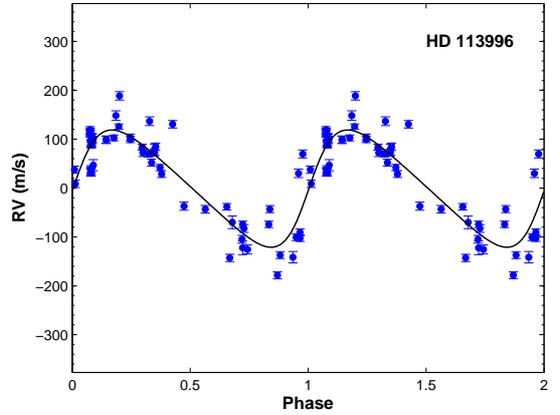}
 \caption{The phase RV curve folded by an orbital period 610 d.
      }
         \label{fig8}
   \end{figure}

 \begin{figure} [h]
 \centering
 \includegraphics[width=8cm]{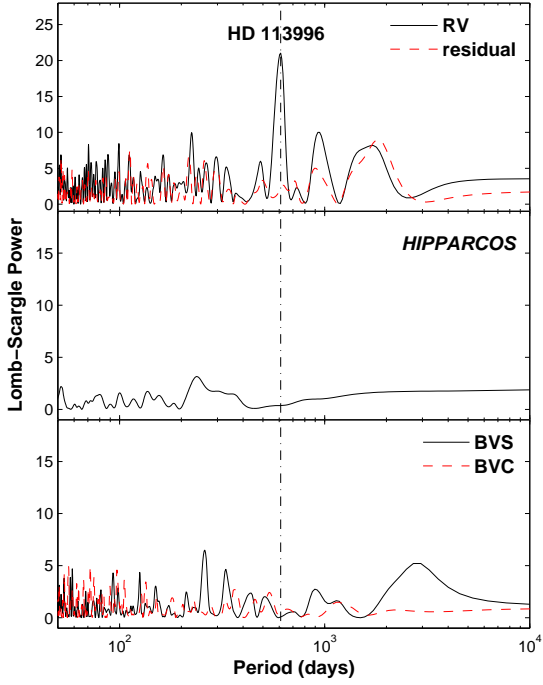}
 \caption{The L-S periodograms of the RVs and residuals (\emph{top}), \emph{HIPPARCOS} photometries (\emph{middle}), and the line bisector span (BVS, solid line) and curvature (BVC, dash line) (\emph{bottom}) for HD 113996. The vertical dash-dot line indicates a period of 610 d.}
 \label{fig9}
 \end{figure}

\section{Stellar activity and pulsation diagnostics}
The RV variations of the stars may be due to some intrinsic nature of the stars, such as stellar pulsation, chromospheric activity, and rotational modulation of surface features. In this Section, we present the investigation of chromospheric activity, photometric, and line shape variation. We also discuss the possibility of pulsation of the stars.

\subsection{Chromospheric activity}
\begin{figure} [h]
 \centering
 \includegraphics[width=8cm]{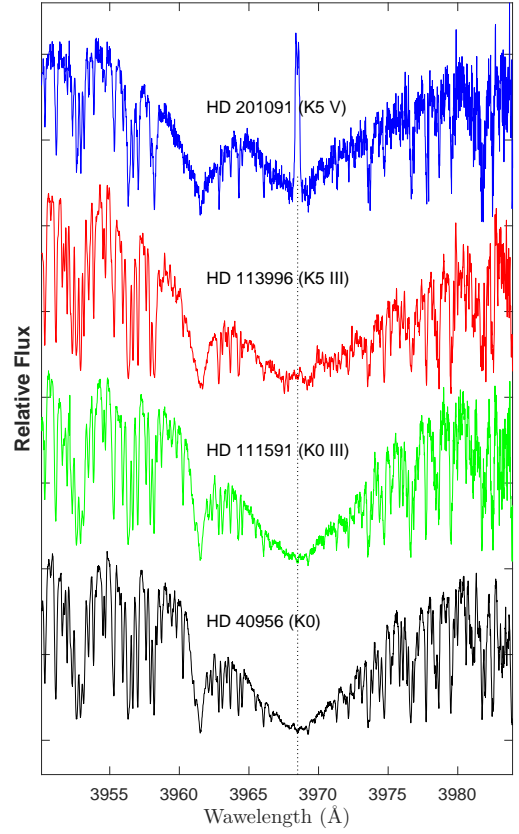}
 \caption{Spectra of Ca II H lines of the HD 40956, HD 111591, and HD 113996. For comparison, we show the spectra of an active K5-type giant star HD 201096 at the top of Figure. The vertical dotted line indicates the center of the Ca II H line (\mbox{3968.5 \AA)}.}
         \label{fig10}
   \end{figure}

To check the stellar activity of the program stars, we show a \mbox{Ca II H} line of each star in Figure \ref{fig10}, which is a sensitive indicator of chromospheric activity. As seen in Figure~\ref{fig10}, we could not find any significant emission in the line core of each star. Although there seems to be a slight reversal in the core of the line of HD 113996, it is negligible compared to the line core of active star HD 201096.

\subsection{HIPPARCOS photometry}
To investigate the photometric variations of the stars, we used \emph{HIPPARCOS} photometric data. From the \emph{HIPPARCOS} catalog, we used 126, 93, and 112 photometric observations for \mbox{HD 40956}, \mbox{HD 111591}, and \mbox{HD 113996}, respectively.
The rms scatters are 0.012, 0.011, and 0.005 mag, respectively.
The rms scatter seems to be comparable to the uncertainty of photometric data which is about 0.005 to 0.008 mag.
We calculated L-S periodogram of the \emph{HIPPARCOS} photometric data, and could not find any periodicity consistent with the RV periods presented in Section 4.

\subsection{Line-shape variation}
 \begin{figure*} [h]
 \centering
 \includegraphics[width=16cm]{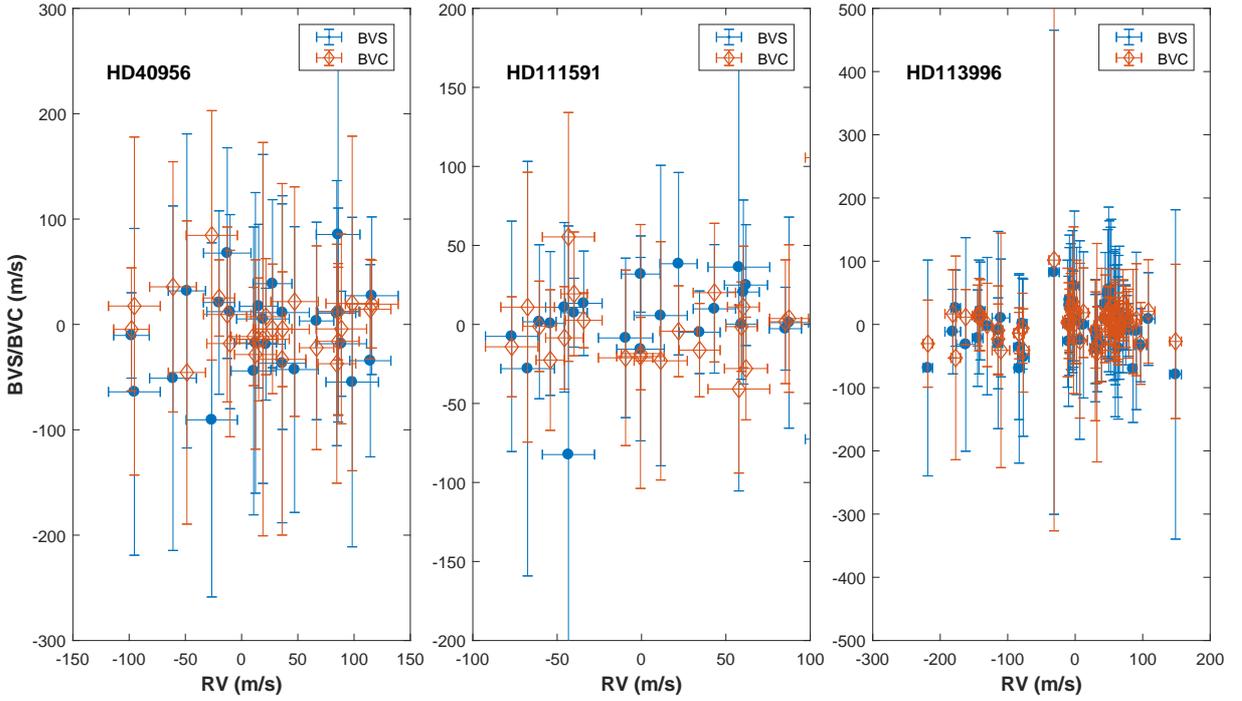}
 \caption{Line bisector measurements versus RV.}
   \label{fig11}
   \end{figure*}
   
Another technique to investigate the cause of the RV variations is to check the line-shape variations. Line-shape variations, pulsations, rotational modulation, and inhomogeneous surface features due to spots or plages. In order to investigate the line-shape variations, we calculated two quantities. One is a bisector velocity span (BVS = V$_{top}$ $-$ V$_{bottom}$), which is a velocity difference between flux levels at 0.4 and 0.8 times the continuum level. The other is a velocity curvature (BVC = [V$_{top}$ $-$ V$_{center}$] $-$ [V$_{center}$ $-$ V$_{bottom}$]), which is the difference between velocity spans in the upper half and the lower half. For the bisector analysis, we selected several lines (\mbox{Ca I} 6122.2, 6439.1, 6462.6, 6717.7; \mbox{Fe I} 6141.7, 6393.6, 6677.9, 6750.2; \mbox{Fe II} 6151.6; \mbox{Ni I} 6643.6, 6767.8; \mbox{Ti I} 6742.6) in the spectral region of 6000 - 7000 \AA, which are relatively strong and free from iodine and telluric lines (\citealt{2005A&A...437..743H,2014JKAS...47...69L,2015A&A...584A..79L}). We performed L-S period analysis for BVS and BVC as shown in the bottom panels of Figures \ref{fig3}, \ref{fig6}, and \ref{fig9} and could not find any periodicity of line-shape variations consistent with those of the RV variations.
To investigate the correlation between RV variations and line bisectors, we plotted the RVs and line bisector values in Figure~\ref{fig11}. 
Although the uncertainties of the line bisector values are large, we cannot see any clear correlation between RV and line bisector.

\subsection{Pulsations}
There are some K giants that are known to be pulsating stars, which have several oscillation modes at different frequencies with different amplitudes.
The pulsation periods can range from hours to days depending on the radius and mass of the star.
For radial pulsations, we can use the Q-constant formulae of \citet{2007MNRAS.378.1270X} to calculate the period, which results in \mbox{P$_{osc}$ $\sim$ 0.58 d} for \mbox{HD 40956}, \mbox{P$_{osc}$ $\sim$ 0.31 d} for \mbox{HD 111591}, and \mbox{P$_{osc}$ $\sim$ 4.78 d} for \mbox{HD 113996}.

The periods for p-mode non-radial pulsations can be estimated from the scaling relationships for $\nu_{max}$, the frequency of maximum pulsation power, which is given by $\nu_{max}$ $\sim$ 3.05 (M/R$^{2}$) mHz. The estimation yields \mbox{P$_{osc}$ $\sim$ 3.3 h} for \mbox{HD 40956}, \mbox{P$_{osc}$ $\sim$ 3 h} for \mbox{HD 111591}, and \mbox{P$_{osc}$ $\sim$ 1.6 d} for \mbox{HD 113996}.

All these periods are much less than the long RV periods found in this study, 
so we can exclude traditional oscillations as the cause of the RV variations.
However some authors report pulsation periods for 
K and M giant stars of  several hundred days;
for example, \citet{2016AJ....151..106L} found a 473 d periodic RV variation for M-giant binary \mbox{$\mu$ UMa}.
They found a correlation between the RV variations and the equivalent widths of H$\alpha$ and H$\beta$ lines, and thus concluded
that the detected RV variations of \mbox{$\mu$ UMa} are caused by non-radial
pulsation. \citet{2015MNRAS.452.3863S} also found possible
convective oscillation modes in three low-mass luminous red giants
in LMC and the modes can be explained by sequence D
pulsations.
Hatzes et al.(2017,  priv. comm.) found long-period variations in $\gamma$ Dra possibly caused by pulsations.
The mechanism of these long-period pulsation of giant stars is not well understood yet.
We therefore reserve the possibility of pulsations as the cause of long-period RV variation of giant stars.

\section{Summary and Discussion}
We clearly detected long-period RV variations in three K giant stars.
In order to confirm the nature of these periodic RV variations, we checked \emph{HIPPARCOS} photometeric data, chromospheric activity, line-shape variations, and rotational period.
These analyses do not show any clear correlation with the RV variations and the rotational periods are much shorter than the observed RV variation periods.
Though we don't exclude pulsation as the cause of the RV variations, as discussed in the previous Section, we conclude that the periodic RV variations are most likely caused by substellar companions.
We performed orbital analysis and found the period and companion mass; P = 578.6 $\pm$ 3.3 d, $m$ sin $i$ = 2.7 $\pm$ 0.6 $M_{\rm{J}}$  for \mbox{HD 40956}; P = 1056.4 $\pm$ 14.3 d, $m$ sin $i$ = 4.4 $\pm$ 0.4 $M_{\rm{J}}$ for \mbox{HD 111591}; P = 610.2 $\pm$ 3.8 d, $m$ sin $i$ = 6.3 $\pm$ 1.0 $M_{\rm{J}}$ for \mbox{HD113996}.

HD 40956 shows complex RV variations with a period of 578.6 d and a long-term linear RV trend of 13 m s$^{-1}$ yr$^{-1}$, which may be related to a long-period companion.
The residuals, after subtracting a 578-day orbital motion, show weak 185-day periodicity.
At this moment,  the reality and cause of this periodicity are not clear. We need more follow-up observations to clarify these issues.
We note that among the planetary systems around giant stars, some systems with two giant planets are found with orbital periods similar to a period ratio of 360 d and 580 d and stable orbits (\citealt{2011AJ....141...16J}, \citealt{2014A&A...568A..64T}, \citealt{2016ApJ...819...59S}).
HD 40956 of this study may add one more example similar to these planetary systems with a period of 185 d instead of 360 d.

HD 111591 has a period of 1056.4 d, which is the longest among the three stars in this study.
Although this long period is comparable to the luminous Mira-type variables,
taking into account the spectral type and surface gravity of this star,
we exclude the Mira-type pulsations as the explanation of the RV variations.
No additional significant periodicity was found in the residual RV after the orbital fit.

The rms residual after the orbital fit of HD 113996 is 39.3 \mbox{m s$^{-1}$}, which is about twice larger than that for the other K0 giants in this study.
As mentioned earlier, it is not surprising that later spectral-type giant stars in general show larger intrinsic RV variations (HD 113996 is K5 III star).
In fact, the typical intrinsic RV variation of K5 stars is larger than 50 m s$^{-1}$;
the K5 III star $\alpha$ Tau, for example, shows about 80 m s$^{-1}$ intrinsic RV variations (\citealt{2015A&A...580A..31H}).

\acknowledgements

This work is supported by the KASI (Korea Astronomy and Space Science Institute) through grant No. 2016-1-860-02. This work was in part supported by JSPS (Japan Society for the Promotion of Science) KAKENHI Grant Numbers 23244038, 16H02169. BCL acknowledges partial support by the KASI grant 2016-1-832-01. This research made use of the SIMBAD database, operated at the CDS, Strasbourg, France. Data analysis was in part carried out on the open use data analysis computer system at the Astronomy Data Center, ADC, of the National Astronomical Observatory of Japan. We thank all officials of BOAO and OAO. We also thank an anonymous referee for helpful comments and suggestions to greatly improve the paper.

\bibliographystyle{apj}
\bibliography{reference}

\begin{thebibliography}{}

\bibitem[\protect\citeauthoryear{{Anderson} \& {Francis}}{{Anderson} \&
  {Francis}}{2012}]{2012AstL...38..331A}
{Anderson}, E.,  \& {Francis}, C. 2012, Astronomy Letters, 38, 331

\bibitem[\protect\citeauthoryear{{Butler} et~al.}{{Butler}
  et~al.}{1996}]{1996PASP..108..500B}
{Butler}, R.~P., {Marcy}, G.~W., {Williams}, E., {McCarthy}, C., {Dosanjh}, P.,
   \& {Vogt}, S.~S. 1996, \pasp, 108, 500

\bibitem[\protect\citeauthoryear{{Collier Cameron} et~al.}{{Collier Cameron}
  et~al.}{2010}]{2010MNRAS.407..507C}
{Collier Cameron}, A., et~al. 2010, \mnras, 407, 507

\bibitem[\protect\citeauthoryear{{da Silva} et~al.}{{da Silva}
  et~al.}{2006}]{2006A&A...458..609D}
{da Silva}, L., et~al. 2006, \aap, 458, 609

\bibitem[\protect\citeauthoryear{{ESA}}{{ESA}}{1997}]{1997yCat.1239....0E}
{ESA}. 1997, VizieR Online Data Catalog, 1239

\bibitem[\protect\citeauthoryear{{Ford} et~al.}{{Ford}
  et~al.}{2011}]{2011ApJS..197....2F}
{Ford}, E.~B., et~al. 2011, \apjs, 197, 2

\bibitem[\protect\citeauthoryear{{Frink} et~al.}{{Frink}
  et~al.}{2002}]{2002ApJ...576..478F}
{Frink}, S., {Mitchell}, D.~S., {Quirrenbach}, A., {Fischer}, D.~A., {Marcy},
  G.~W.,  \& {Butler}, R.~P. 2002, \apj, 576, 478

\bibitem[\protect\citeauthoryear{{Gaudi} et~al.}{{Gaudi}
  et~al.}{2017}]{2017Natur.546..514G}
{Gaudi}, B.~S., et~al. 2017, \nat, 546, 514

\bibitem[\protect\citeauthoryear{{Han} et~al.}{{Han}
  et~al.}{2007}]{2007PKAS...22...75H}
{Han}, I., {Kim}, K.-M., {Byeong-Cheol}, L.,  \& {Valyavin}, G. 2007,
  Publication of Korean Astronomical Society, 22

\bibitem[\protect\citeauthoryear{{Han} et~al.}{{Han}
  et~al.}{2008}]{2008JKAS...41...59H}
{Han}, I., {Lee}, B.-C., {Kim}, K.-M.,  \& {Mkrtichian}, D.~E. 2008, Journal of
  Korean Astronomical Society, 41, 59

\bibitem[\protect\citeauthoryear{{Han} et~al.}{{Han}
  et~al.}{2010}]{2010A&A...509A..24H}
{Han}, I., {Lee}, B.~C., {Kim}, K.~M., {Mkrtichian}, D.~E., {Hatzes}, A.~P.,
  \& {Valyavin}, G. 2010, \aap, 509, A24

\bibitem[\protect\citeauthoryear{{Hartman} et~al.}{{Hartman}
  et~al.}{2015}]{2015AJ....150..197H}
{Hartman}, J.~D., et~al. 2015, \aj, 150, 197

\bibitem[\protect\citeauthoryear{{Hatzes} \& {Cochran}}{{Hatzes} \&
  {Cochran}}{1993}]{1993ApJ...413..339H}
{Hatzes}, A.~P.,  \& {Cochran}, W.~D. 1993, \apj, 413, 339

\bibitem[\protect\citeauthoryear{{Hatzes} et~al.}{{Hatzes}
  et~al.}{2015}]{2015A&A...580A..31H}
{Hatzes}, A.~P., et~al. 2015, \aap, 580, A31

\bibitem[\protect\citeauthoryear{{Hatzes} et~al.}{{Hatzes}
  et~al.}{2005}]{2005A&A...437..743H}
{Hatzes}, A.~P., {Guenther}, E.~W., {Endl}, M., {Cochran}, W.~D.,
  {D{\"o}llinger}, M.~P.,  \& {Bedalov}, A. 2005, \aap, 437, 743

\bibitem[\protect\citeauthoryear{{Hekker} et~al.}{{Hekker}
  et~al.}{2006}]{2006A&A...454..943H}
{Hekker}, S., {Reffert}, S., {Quirrenbach}, A., {Mitchell}, D.~S., {Fischer},
  D.~A., {Marcy}, G.~W.,  \& {Butler}, R.~P. 2006, \aap, 454, 943

\bibitem[\protect\citeauthoryear{{Izumiura}}{{Izumiura}}{1999}]{1999PYunO....S..77I}
{Izumiura}, H. 1999, Publications of the Yunnan Observatory, 77

\bibitem[\protect\citeauthoryear{{Izumiura}}{{Izumiura}}{2005}]{2005JKAS...38...81I}
{Izumiura}, H. 2005, Journal of Korean Astronomical Society, 38, 81

\bibitem[\protect\citeauthoryear{{Johnson} et~al.}{{Johnson}
  et~al.}{2011}]{2011AJ....141...16J}
{Johnson}, J.~A., et~al. 2011, \aj, 141, 16

\bibitem[\protect\citeauthoryear{{Kambe} et~al.}{{Kambe}
  et~al.}{2002}]{2002PASJ...54..865K}
{Kambe}, E., et~al. 2002, \pasj, 54, 865

\bibitem[\protect\citeauthoryear{{Kambe} et~al.}{{Kambe}
  et~al.}{2013}]{2013PASJ...65...15K}
{Kambe}, E., et~al. 2013, \pasj, 65, 15

\bibitem[\protect\citeauthoryear{{Kim} et~al.}{{Kim}
  et~al.}{2007}]{2007PASP..119.1052K}
{Kim}, K.-M., et~al. 2007, \pasp, 119, 1052

\bibitem[\protect\citeauthoryear{{Lee}, {Han}, \& {Park}}{{Lee}
  et~al.}{2013}]{2013A&A...549A...2L}
{Lee}, B.-C., {Han}, I.,  \& {Park}, M.-G. 2013, \aap, 549, A2

\bibitem[\protect\citeauthoryear{{Lee} et~al.}{{Lee}
  et~al.}{2016}]{2016AJ....151..106L}
{Lee}, B.-C., {Han}, I., {Park}, M.-G., {Mkrtichian}, D.~E., {Hatzes}, A.~P.,
  {Jeong}, G.,  \& {Kim}, K.-M. 2016, \aj, 151, 106

\bibitem[\protect\citeauthoryear{{Lee} et~al.}{{Lee}
  et~al.}{2014a}]{2014A&A...566A..67L}
{Lee}, B.-C., {Han}, I., {Park}, M.-G., {Mkrtichian}, D.~E., {Hatzes}, A.~P.,
  \& {Kim}, K.-M. 2014a, \aap, 566, A67

\bibitem[\protect\citeauthoryear{{Lee} et~al.}{{Lee}
  et~al.}{2014b}]{2014JKAS...47...69L}
{Lee}, B.-C., {Han}, I., {Park}, M.-G., {Mkrtichian}, D.~E., {Jeong}, G.,
  {Kim}, K.-M.,  \& {Valyavin}, G. 2014b, Journal of Korean Astronomical
  Society, 47, 69

\bibitem[\protect\citeauthoryear{{Lee} et~al.}{{Lee}
  et~al.}{2015}]{2015A&A...584A..79L}
{Lee}, B.-C., et~al. 2015, \aap, 584, A79

\bibitem[\protect\citeauthoryear{{Omiya} et~al.}{{Omiya}
  et~al.}{2012}]{2012PASJ...64...34O}
{Omiya}, M., et~al. 2012, \pasj, 64, 34

\bibitem[\protect\citeauthoryear{{Omiya} et~al.}{{Omiya}
  et~al.}{2009}]{2009PASJ...61..825O}
{Omiya}, M., et~al. 2009, \pasj, 61, 825

\bibitem[\protect\citeauthoryear{{Reffert} et~al.}{{Reffert}
  et~al.}{2015}]{2015A&A...574A.116R}
{Reffert}, S., {Bergmann}, C., {Quirrenbach}, A., {Trifonov}, T.,  \&
  {K{\"u}nstler}, A. 2015, \aap, 574, A116

\bibitem[\protect\citeauthoryear{{Saio} et~al.}{{Saio}
  et~al.}{2015}]{2015MNRAS.452.3863S}
{Saio}, H., {Wood}, P.~R., {Takayama}, M.,  \& {Ita}, Y. 2015, \mnras, 452,
  3863

\bibitem[\protect\citeauthoryear{{Sato} et~al.}{{Sato}
  et~al.}{2007}]{2007ApJ...661..527S}
{Sato}, B., et~al. 2007, \apj, 661, 527

\bibitem[\protect\citeauthoryear{{Sato} et~al.}{{Sato}
  et~al.}{2002}]{2002PASJ...54..873S}
{Sato}, B., {Kambe}, E., {Takeda}, Y., {Izumiura}, H.,  \& {Ando}, H. 2002,
  \pasj, 54, 873

\bibitem[\protect\citeauthoryear{{Sato} et~al.}{{Sato}
  et~al.}{2005}]{2005PASJ...57...97S}
{Sato}, B., {Kambe}, E., {Takeda}, Y., {Izumiura}, H., {Masuda}, S.,  \&
  {Ando}, H. 2005, \pasj, 57, 97

\bibitem[\protect\citeauthoryear{{Sato} et~al.}{{Sato}
  et~al.}{2012}]{2012PASJ...64..135S}
{Sato}, B., et~al. 2012, \pasj, 64, 135

\bibitem[\protect\citeauthoryear{{Sato} et~al.}{{Sato}
  et~al.}{2016}]{2016ApJ...819...59S}
{Sato}, B., et~al. 2016, \apj, 819, 59

\bibitem[\protect\citeauthoryear{{Takeda} et~al.}{{Takeda}
  et~al.}{2005}]{2005PASJ...57...27T}
{Takeda}, Y., {Ohkubo}, M., {Sato}, B., {Kambe}, E.,  \& {Sadakane}, K. 2005,
  \pasj, 57, 27

\bibitem[\protect\citeauthoryear{{Takeda}, {Sato}, \& {Murata}}{{Takeda}
  et~al.}{2008}]{2008PASJ...60..781T}
{Takeda}, Y., {Sato}, B.,  \& {Murata}, D. 2008, \pasj, 60, 781

\bibitem[\protect\citeauthoryear{{Trifonov} et~al.}{{Trifonov}
  et~al.}{2014}]{2014A&A...568A..64T}
{Trifonov}, T., {Reffert}, S., {Tan}, X., {Lee}, M.~H.,  \& {Quirrenbach}, A.
  2014, \aap, 568, A64

\bibitem[\protect\citeauthoryear{{van Leeuwen}}{{van
  Leeuwen}}{2007}]{2007A&A...474..653V}
{van Leeuwen}, F. 2007, \aap, 474, 653

\bibitem[\protect\citeauthoryear{{Xiong} \& {Deng}}{{Xiong} \&
  {Deng}}{2007}]{2007MNRAS.378.1270X}
{Xiong}, D.~R.,  \& {Deng}, L. 2007, \mnras, 378, 1270

\bibitem[\protect\citeauthoryear{{Zhou} et~al.}{{Zhou}
  et~al.}{2017}]{2017AJ....153..211Z}
{Zhou}, G., et~al. 2017, \aj, 153, 211

\bibitem[\protect\citeauthoryear{{Zhou} et~al.}{{Zhou}
  et~al.}{2016}]{2016AJ....152..136Z}
{Zhou}, G., et~al. 2016, \aj, 152, 136

\end{thebibliography}

\clearpage

\begin{table}
\begin{center}
\caption[]{RV measurements for HD 40956}
\label{tab3}
\begin{tabular}{cccc}
\toprule[1.5pt]
    JD       & $\Delta$RV   & $\pm$$\sigma$ & \multirow{2}*{Observatory / Spectrograph} \\
-2450000  & m s$^{-1}$ & m s$^{-1}$ &\\
\midrule
  5323.9789 & 15.2 & 17.0 & BOAO / BOES\\
  5527.0623 & 21.6 & 17.2 & BOAO / BOES\\
  5553.2149 & 114.4 & 18.3 & BOAO / BOES\\
  5583.1940 & 88.7 & 22.5 & BOAO / BOES\\
  5636.9949 & 85.8 & 19.4 & BOAO / BOES\\
  5881.3676& -18.7 & 3.9 & OAO / HIDES slit\\
  5912.2101 & -12.7 & 21.1 & BOAO / BOES\\
  5917.1522 & -37.2 & 7.4 & OAO / HIDES slit\\
  5920.1383 & -31.4 & 4.7 & OAO / HIDES slit\\
  5939.1603 & -13.8 & 4.9 & OAO / HIDES slit\\
  5968.0579 & -10.3 & 20.7 & BOAO / BOES\\
  5974.0024 & -12.6 & 4.6 & OAO / HIDES slit\\
  6010.9628 & -31.9 & 4.8 & OAO / HIDES slit\\
  6031.9598 & -29.6 & 3.8 & OAO / HIDES slit\\
  6212.3446 & 48.9 & 4.7 & OAO / HIDES slit\\
  6235.3600 & 80.5 & 4.5 & OAO / HIDES slit\\
  6238.2177 & 73.9 & 6.4 & OAO / HIDES slit\\
  6250.1645 & 87.0 & 5.2 & OAO / HIDES slit\\
  6280.1870 & 115.7 & 23.5 & BOAO / BOES\\
  6286.1746 & 109.8 & 4.0 & OAO / HIDES slit\\
  6332.0850 & 85.9 & 18.6 & BOAO / BOES\\
  6385.9881& 36.3 & 23.9 & BOAO / BOES\\
  6515.3211 & -69.7 & 4.1 & OAO / HIDES fiber\\
  6553.3025 & -97.9 & 15.8 & BOAO / BOES\\
  6554.2186 & -59.0 & 3.7 & OAO / HIDES fiber\\
  6556.3094 & -48.6 & 16.5 & BOAO / BOES\\
  6576.2283& -41.2 & 7.9 & OAO / HIDES slit\\
  6580.2782 & -31.8 & 4.9 & OAO / HIDES slit\\
  6595.2285 & -20.1 & 3.8 & OAO / HIDES fiber\\
  6607.1773 & -5.2 & 3.8 & OAO / HIDES fiber\\
  6613.0858 & 10.7 & 20.1 & BOAO / BOES\\
  6615.3592 & -2.6 & 4.1 & OAO/HIDES fiber\\
  6625.1445 & 7.9 & 3.6 & OAO / HIDES fiber\\
  6639.1269& 21.0 & 3.9 & OAO / HIDES fiber\\
  6658.0666 & 24.8 & 4.0 & OAO / HIDES fiber\\
  6661.0665 & 29.1 & 3.7 & OAO / HIDES fiber\\
  6667.0308 & 13.0 & 5.2 & OAO / HIDES slit\\
  6675.0717 & 36.0 & 21.1 & BOAO / BOES\\
  6680.0901 & 39.4 & 3.7 & OAO / HIDES fiber\\
  6697.9855& 41.2 & 4.0 & OAO / HIDES fiber\\
  6713.9426& 35.6 & 3.0 & OAO / HIDES fiber\\
  6732.9673 & 33.6 & 3.9 & OAO / HIDES fiber\\
  6740.9561 & 29.2 & 4.2 & OAO / HIDES slit\\
  6761.9766 & 45.0 & 3.8 & OAO / HIDES fiber\\
  6964.2531 & 23.8 & 4.0 & OAO / HIDES fiber\\
  7005.1077 & -15.5 & 4.3 & OAO / HIDES fiber\\
  7027.3119 & 19.0 & 23.4 & BOAO / BOES\\
  7031.1283 & -38.5 & 3.9 & OAO / HIDES slit\\
  7067.2337 & -26.5 & 22.7 & BOAO / BOES\\
  7068.1621 & -61.0 & 20.8 & BOAO / BOES\\
  7092.0618 & -95.3 & 23.0 & BOAO / BOES\\
  7092.9540& -87.9 & 3.4 & OAO / HIDES fiber\\
  7108.0408& -75.7 & 3.1 & OAO / HIDES fiber\\
  7141.9433 & -114.5 & 4.2 & OAO / HIDES fiber\\
\bottomrule[1.5pt]

\end{tabular}
\end{center}
\end{table}

\begin{table}
\begin{center}
\caption[]{RV measurements for HD 111591}
\label{tab4}
\begin{tabular}{cccc}
\toprule[1.5pt]
    JD       & $\Delta$RV   & $\pm$$\sigma$ & \multirow{2}*{Observatory/Spectrograph} \\
-2450000  & m s$^{-1}$ & m s$^{-1}$ &\\
\midrule
3427.3029 & 59.3 & 9.3 & BOAO / BOES\\
3756.3732 & -0.5 & 11.5 & BOAO / BOES\\
4834.3355 & -40.0 & 7.8 & BOAO / BOES\\
4931.0714 & -54.1 & 8.4 & BOAO / BOES\\
5248.3155 & 85.3 & 9.5 & BOAO / BOES\\
5324.1067 & 60.4 & 9.4 & BOAO / BOES\\
5358.0604 & 108.1 & 11.0 & BOAO / BOES\\
5583.3769 & 11.4 & 15.8 & BOAO / BOES\\
5639.1594 & -0.6 & 14.1 & BOAO / BOES\\
5705.0934 & -9.4 & 16.1 & BOAO / BOES\\
5912.3642 & -60.5 & 9.9 & BOAO / BOES\\
5969.3377 & -76.9 & 15.6 & BOAO / BOES\\
6070.1044 & -16.3 & 3.8 & OAO / HIDES slit\\
6280.3748 & 87.4 & 11.5 & BOAO / BOES\\
6334.3126 & 61.9 & 12.8 & BOAO / BOES\\
6384.1195 & 57.7 & 18.3 & BOAO / BOES\\
6387.0977 & 34.4 & 12.2 & BOAO / BOES\\
6424.1126 & 47.5 & 4.7 & OAO / HIDES slit\\
6451.0005 & 56.2 & 3.4 & OAO / HIDES slit\\
6514.9560 & 92.2 & 3.7 & OAO / HIDES fiber\\
6625.3257 & 34.8 & 3.9 & OAO / HIDES fiber\\
6660.3878 & 25.2 & 3.4 & OAO / HIDES fiber\\
6675.3470 & 43.2 & 12.6 & BOAO / BOES\\
6676.3999 & 22.0 & 11.0 & BOAO / BOES\\
6680.2737 & 24.8 & 3.6 & OAO / HIDES fiber\\
6733.0800 & -14.8 & 3.5 & OAO / HIDES fiber\\
6741.1461 & -26.3 & 4.3 & OAO / HIDES slit\\
6762.1124 & -33.1 & 3.6 & OAO / HIDES fiber\\
6781.1942 & -34.3 & 10.9 & BOAO / BOES\\
6784.1022 & -40.9 & 3.0 & OAO / HIDES fiber\\
6784.1229 & -45.7 & 11.2 & BOAO / BOES\\
6793.1269 & -26.6 & 3.6 & OAO / HIDES fiber\\
6812.1034 & -29.7 & 4.7 & OAO / HIDES fiber\\
6965.3508 & -29.9 & 4.5 & OAO / HIDES fiber\\
7005.2581 & -12.9 & 3.6 & OAO / HIDES fiber\\
7027.2939 & -43.3 & 15.4 & BOAO / BOES\\
7031.2874 & -25.5 & 3.6 & OAO / HIDES slit\\
7067.2528 & -67.5 & 15.9 & BOAO / BOES\\
7092.0811 & -53.2 & 15.2 & BOAO / BOES\\
7094.1016 & -31.1 & 3.5 & OAO / HIDES fiber\\
7108.1564 & -14.6 & 3.3 & OAO / HIDES fiber\\
7169.0409 & -3.8 & 10.9 & BOAO / BOES\\
7170.0352 & 3.6 & 3.4 & OAO / HIDES fiber\\
7222.9709 & 29.6 & 4.0 & OAO / HIDES fiber\\
\bottomrule[1.5pt]

\end{tabular}
\end{center}
\end{table}

\begin{table}
\begin{center}
\caption[]{RV measurements for HD 113996}
\label{tab5}
\begin{tabular}{cccc}
\toprule[1.5pt]
    JD       & $\Delta$RV   & $\pm$$\sigma$ & \multirow{2}*{Observatory / Spectrograph} \\
-2450000  & m s$^{-1}$ & m s$^{-1}$ &\\
\midrule
3430.1733 & 93.3 & 5.4 & BOAO / BOES\\
3460.2060 & 67.0 & 6.3 & BOAO / BOES\\
3460.2561 & 71.3 & 6.8 & BOAO / BOES\\
3507.1163 & 37.6 & 6.3 & BOAO / BOES\\
3509.9766 & 104.5 & 8.9 & BOAO / BOES\\
3818.1061 & -106.6 & 6.5 & BOAO / BOES\\
3821.0853 & -75.4 & 7.3 & BOAO / BOES\\
3889.1274 & -132.8 & 7.0 & BOAO / BOES\\
3899.0409 & -134.9 & 6.8 & BOAO / BOES\\
3899.1054 & -122.8 & 6.9 & BOAO / BOES\\
4125.2073 & 19.8 & 7.2 & BOAO / BOES\\
4147.2649 & 9.6 & 6.4 & BOAO / BOES\\
4151.3333 & -3.7 & 7.0 & BOAO / BOES\\
4209.1005 & -69.4 & 8.4 & BOAO / BOES\\
4264.1103 & -75.6 & 7.7 & BOAO / BOES\\
4505.3163 & -2.2 & 8.8 & BOAO / BOES\\
4516.2931 & 37.4 & 8.2 & BOAO / BOES\\
4536.1626 & 5.2 & 7.3 & BOAO / BOES\\
4538.1339 & -23.7 & 7.6 & BOAO / BOES\\
4618.0332 & 67.2 & 7.1 & BOAO / BOES\\
4618.0405 & 66.0 & 7.3 & BOAO / BOES\\
4643.1020 & 116.0 & 9.8 & BOAO / BOES\\
4930.0810 & -70.6 & 6.6 & BOAO / BOES\\
4971.0880 & -106.6 & 6.5 & BOAO / BOES\\
5248.1945 & 70.3 & 6.0 & BOAO / BOES\\
5321.0988 & 51.5 & 6.0 & BOAO / BOES\\
5356.1121 & 52.2 & 7.6 & BOAO / BOES\\
5554.3842 & -102.5 & 13.0 & BOAO / BOES\\
5581.1788 & -154.6 & 14.0 & BOAO / BOES\\
5671.2112 & -210.7 & 6.9 & BOAO / BOES\\
5711.1867 & -173.8 & 11.6 & BOAO / BOES\\
5933.3180 & 39.9 & 7.8 & BOAO / BOES\\
5963.2053 & 40.4 & 7.2 & BOAO / BOES\\
5963.3409 & 46.6 & 10.9 & BOAO / BOES\\
6288.3860 & -169.5 & 7.0 & BOAO / BOES\\
6407.0292 & 82.5 & 6.3 & BOAO / BOES\\
6407.0368 & 87.0 & 6.4 & BOAO / BOES\\
6407.0442 & 84.2 & 5.9 & BOAO / BOES\\
6407.0517 & 79.2 & 6.2 & BOAO / BOES\\
6407.0591 & 78.5 & 6.7 & BOAO / BOES\\
6409.0348 & -0.8 & 7.2 & BOAO / BOES\\
6409.0422 & 6.4 & 7.1 & BOAO / BOES\\
6409.0497 & 3.6 & 6.6 & BOAO / BOES\\
6409.0571 & -1.0 & 7.0 & BOAO / BOES\\
6409.0646 & 1.2 & 7.4 & BOAO / BOES\\
6410.0079 & 58.3 & 6.3 & BOAO / BOES\\
6410.0154 & 57.2 & 5.5 & BOAO / BOES\\
6410.0228 & 60.2 & 5.9 & BOAO / BOES\\
6410.0302 & 60.9 & 6.2 & BOAO / BOES\\
6410.0376 & 55.9 & 6.4 & BOAO / BOES\\
6412.1810 & 73.2 & 6.0 & BOAO / BOES\\
6412.1954 & 71.9 & 5.9 & BOAO / BOES\\
6412.2098 & 66.9 & 6.6 & BOAO / BOES\\
6412.2242 & 64.4 & 5.9 & BOAO / BOES\\
6412.2386 & 67.1 & 6.1 & BOAO / BOES\\
6620.3166 & 98.2 & 7.7 & BOAO / BOES\\
6800.1058 & -137.0 & 7.9 & BOAO / BOES\\
7025.2713 & 14.5 & 11.4 & BOAO / BOES\\
7093.0806 & 156.3 & 8.6 & BOAO / BOES\\
7378.3318 & -175.2 & 7.4 & BOAO / BOES\\
7415.1456 & -115.0 & 8.0 & BOAO / BOES\\
7424.2282 & -157.7 & 9.0 & BOAO / BOES\\
\bottomrule[1.5pt]

\end{tabular}
\end{center}
\end{table}

\end{document}